\documentclass[aps,prd,amssymb,eqsecnum,amsfonts,epsf,preprintnumbers,nofootinbib,superscriptaddress]{revtex4}
\usepackage{cases}
\usepackage{graphicx,subfigure}% Include figure files
\usepackage{dcolumn,color}% Align table columns on decimal point
\usepackage{bm}
\usepackage{epstopdf}
\usepackage{epsfig}
\usepackage{color,framed}
\usepackage{amsthm}

\usepackage[table]{xcolor}

\newcommand{\pt}{\partial}

\newcommand{\be}{\begin{equation}}
\newcommand{\ee}{\end{equation}}
\newcommand{\ba}{\begin{eqnarray}}
\newcommand{\ea}{\end{eqnarray}}

\begin{document}

\title{Stability of braneworlds with non-minimally coupled multi-scalar fields}

\author{Feng-Wei Chen \footnote{chenfw10@lzu.edu.cn},
        Bao-Min Gu \footnote{gubm15@lzu.edu.cn}, and
        Yu-Xiao Liu\footnote{liuyx@lzu.edu.cn, corresponding author}
        }
\affiliation{Research Center of Gravitation, Lanzhou University, \\
Lanzhou 730000, China\\
Institute of Theoretical Physics, Lanzhou University,
Lanzhou 730000, China and\\
Key Laboratory for Magnetism and Magnetic Materials of the Ministry of Education, Lanzhou University,
Lanzhou 730000, China}

%\emailAdd{chenfw10@lzu.edu.cn, gubm15@lzu.edu.cn, and liuyx@lzu.edu.cn}

\begin{abstract}{ Linear stability of braneworld models constructed with multi-scalar fields is very different from that of single-scalar field models. It is well known that both the tensor and the scalar perturbations  of the latter
are stable at linear level. However, in general
there is no effective method to deal with the stability problem of the scalar perturbations for braneworld models constructed with non-minimally coupled multi-scalar fields. In this work we present a systematic covariant approach to deal with the scalar perturbations. By introducing the orthonormal bases in field space and making the Kaluza--Klein decomposition, we get a set of coupled Schr\"{o}dinger-like equations of the scalar perturbation modes. Using the nodal theorem, we show that the result is model-dependent. For superpotential derived brane models, the scalar perturbations are stable, but there exist normalizable scalar zero modes, which will result in unacceptable fifth force on the brane. We also use this method to analyze the $f(R)$ braneworld model with an explicit solution and find that the scalar perturbations are stable and the scalar zero modes cannot be localized on the brane, which ensures that there is no extra long-range force and the Newtonian potential on the brane can be recovered. }
\end{abstract}
\maketitle
%\tableofcontents

\section{Introduction}

%Braneworld opened up a possible way to find new physics beyond the Standard Model \cite{Arkani-HamedDimopoulosDvaliMarch-Russell2002, RandallSundrum1999a, RandallSundrum1999}. The most famous braneworld models are Randall-Sundrum (RS) models. In the original RS-I model, there are two branes without  thickness or inner structure (called thin branes), and there is no scalar field in the bulk. However, to stabilize the distance between two thin branes in RS-I \cite{Goldberger:1999uk,Goldberger:1999un} and to obtain smooth version of the RS-II \cite{DeWolfeFreedmanGubserKarch2000,DzhunushalievFolomeevMinamitsuji2010}, scalar fields are needed.
%In the simple case, there is a single scalar which interpolates two vacua. Multiple scalar models were also studied in Refs. \cite{Eto:2003bn,Bazeia:2004dh}.
%

The braneworld scenario has opened up a new way to search for new physics beyond the Standard Model \cite{Arkani-Hamed2002,RandallSundrum1999,RandallSundrum1999a}. The most famous braneworld models are the Randall--Sundrum (RS) models \cite{RandallSundrum1999,RandallSundrum1999a}, which were proposed to solve the gauge hierarchy problem. In the original RS-I/II model, there are two/one branes   without thickness or inner structure (called thin branes). There is no background scalar field in the bulk and the $AdS_5$ geometry is obtained by adding a negative cosmological constant in the bulk and a brane tension on each brane. In the RS-II model \cite{RandallSundrum1999a}, a remarkable discovery is that the four-dimensional gravity (the Newtonian potential) can be recovered on the brane even though the extra dimension is infinite \cite{RandallSundrum1999a,Giddings:2000mu,PhysRevLett.84.2778}.

However, to get the smooth version of the RS-II model, scalar fields as material are introduced~\cite{DeWolfe2000}. On the other hand, in order to stabilize the size of the extra dimension in the RS scenario, a bulk scalar is also necessary~\cite{GoldbergerWise1999}.
The most intuitive idea is to consider a canonical scalar field. As was proposed in Ref.~\cite{Rubakov:1983bb}, a kink scalar field can be used to construct a domain wall configuration, so that ordinary particles can be confined in a potential well. If gravity is included, then an RS-II like braneworld model can be obtained, with a smooth asymptotically $AdS_5$ geometry. Because of the domain wall configuration, the four-dimensional massless graviton is trapped in an effective potential well while the massless longitudinal mode (scalar zero mode) cannot be localized, thus the four-dimensional gravity can also be recovered in such models \cite{Gremm:1999pj,Csaki:2000fc,ThickBrane2002}. The localization of standard model fields relies on different mechanisms, namely, some special couplings \cite{Oda:2000zc,Ringeval:2001cq,Liu:2013kxz,LiLiu2017a}.

There is extended work in which multiple canonical scalar fields were considered. The situation will become completely different. This setup would lead to some interesting internal structures for the brane \cite{Campos:2001pr,Eto:2003bn,Bazeia:2004dh} and special properties of localization of matter fields \cite{Almeida2009,LiuLi2009aa,Correa1011.1849,Xie2013}.
However, it is very important to ensure that the linear perturbations are stable and the scalar zero modes should not be localized on the brane in order to recover the right effective four-dimensional gravity.
The stability of the scalar perturbations for such brane was studied in Refs.~\cite{Aybat2010,George2011,Gherghetta:2011rr}. In fact, the special model with two scalars is the so called Bloch brane \cite{Bazeia:2004dh}.  However, in Ref.~\cite{Aybat2010}, it was shown that in such models only odd scalars can avoid the existence of scalar zero modes. Indeed, for models constructed with two scalar fields, if the background solutions are derived from superpotential, then a localized scalar zero mode would appear inevitably \cite{George2011}. As is well known, a localized scalar zero mode would lead to an extra long-range force that has never been observed, and hence it is unacceptable.

In addition to the canonical fields, non-canonical structures are also considered because of their special dynamics. For example, the Born--Infeld type matter action was widely studied in the literature \cite{Felder:2002sv,Alishahiha:2004eh}. In addition, K-fields also aroused great interests of cosmologists since this kind of non-canonical scalar field is believed to be able to drive inflation with generic initial conditions \cite{Armendariz1999rj,Garriga1999vw}. The domain wall brane models constructed with K-field were proved to be stable under scalar perturbations \cite{Adam:2007ag,Bazeia:2008zx,Zhong:2012nt,Zhong:2013xga}. This gives us inspiration that the localized zero modes in braneworld model with multiple canonical scalars may be avoided if the scalars are non-canonical. However, we will not use K-fields. Instead, we will investigate  braneworlds generated by non-minimally coupled multi-scalar fields, which would be non-canonical in the Einstein frame.
We are motivated to consider such models from different aspects. The first and natural reason comes from string theory. It has been shown that in low energy limit, the bosonic string theory reduces to scalar--tensor theory, not general relativity. The study of braneworld models in scalar--tensor theory gives a lower -dimensional understanding of string theory. Besides, it is widely believed that the multiple fields models would give very interesting braneworld structures \cite{Bazeia:2004dh,Almeida:2009jc,Zhao:2009ja,BazeiaMenezesPetrovSilva2013}, such as the brane splitting and the gravity resonance.
In Ref.~\cite{Liu:2012gv}, the domain wall brane constructed with a single non-minimally coupled scalar field was studied. The analysis on the full linear perturbations shows that the massless graviton can be trapped on the brane, and the bound state of the scalar perturbation mode should not exist, but it still needs a rigorous proof.
We expect to give a systematic research on scalar perturbations of braneworlds with non-minimally coupled multiple scalars, including their stability and localization properties.

Our research is meaningful not only for the above-mentioned braneworld models themselves, but also for other models such as the widely studied $f(R)$ braneworld model~\cite{ParryPichlerDeeg2005,AfonsoBazeiaMenezesPetrov2007,DeruelleSasakiSendouda2008,BalcerzakDabrowski2010,Bouhmadi-LopezCapozzielloCardone2010,DzhunushalievFolomeevKleihausKunz2010,HoffDias2011,LiuZhongZhaoLi2011,LiuLuWang2012,BazeiaMenezesPetrovSilva2013,BazeiaLobaoMenezesPetrovSilva2014,ZhongLiuYang2011,XuZhongYuLiu2015,YuZhongGuLiu2016}, in which the scalar perturbations are still not clear because of the higher derivatives in the perturbation equations. It is well known that the $f(R)$ gravity theory is equivalent to the scalar--tensor theory, and both the non-minimally coupling gravity theory and $f(R)$ gravity theory can be cast in terms of the Einstein frame but with non-canonical scalar fields (see Sec. \ref{secfR} for details). Hence a natural application of our results is the $f(R)$ braneworld model.

In this paper we choose a general action and mainly study the scalar perturbations of braneworlds generated by non-minimally coupled multi-scalar fields in the Einstein frame, by using some techniques developed in cosmology.
The paper is organized as follows. In the following section, we briefly introduce our model and derive
the perturbation equations of the scalar modes. In Sec. \ref{zero}, we investigate the perturbation equations and analyze the stability of the massless scalar modes. We deal with the scalar perturbations of the $f(R)$ braneworld model in Sec. \ref{secfR} and give a summary in the last section.

{\section{The model}}
{\subsection{General setup}}
Following Ref. \cite{Gong2011}, we adopt the action
\begin{eqnarray}
S=\int \mathrm{d}^nx\sqrt{-g}\left[\frac{1}{2\kappa_n}R
 +P\left(\mathcal{G}_{IJ},X^{IJ},f_a^{J_1\cdots J_{n_a}}(\Phi^I)\right)\right],
\end{eqnarray}
where $\kappa_n=8\pi G_n$ with $G_n$ the $n$-dimensional Newtonian constant and will be set to be 1 ($\kappa_n=1$) in this paper for simplicity, $X^{IJ}=-\frac{1}{2}\partial_M\Phi^I\partial^M\Phi^J$ is the kinetic function,  $\mathcal{G}^{IJ}=\mathcal{G}^{IJ}(\phi^K)$ can be interpreted as the metric on the field space,\footnote{In the following, indices $I,J,K,L,R,S$ denote field-space indices, which are lowered or raised by the field-space metric $\mathcal{G}$ or its inverse, while $M,N,P,Q$ etc run over $n$-dimensional ones of the spacetime.} and $f_a^{J_1\cdots J_{n_a}}(\Phi^I)$ are some field-space tensors with the subscript $a$ introduced to discriminate different kinds of such tensors. Here we assume that $P$ depends on the scalar fields only though field-space tensors $f_a^{J_1\cdots J_{n_a}}$. We also assume that there is no spacetime derivatives of fields in $f_a^{J_1\cdots J_{n_a}}$.  The most important and simplest case is the potential $V(\Phi^I)$, which is a field-space scalar. In this paper, we assume that the field-space tensors are all scalars for simplicity, and hence
\begin{eqnarray}
P=P(\mathcal{G}_{IJ},X^{IJ},f_a(\Phi^I)).
\end{eqnarray}
If we can choose such coordinates $\{\tilde{\Phi}\}$ to make the field-space metric $\mathcal{G}$ trivial, i.e.
\begin{eqnarray}
\mathcal{G}_{IJ} \frac{\partial\tilde{\Phi}^I}{\partial\Phi^K}
           \frac{\partial\tilde{\Phi}^J}{\partial\Phi^L} =\delta_{KL},
\end{eqnarray}
then the corresponding curvature $\mathcal{R}_{IJKL}$ constructed from the field-space metric $\mathcal{G}_{IJ}$ vanishes.
The simplest case is
\be
P=\mathcal{G}_{IJ}X^{IJ}-V(\Phi).
\ee
When $\mathcal{G}_{IJ}=\delta_{IJ}$, all scalar fields are canonical.
The ansatz of the background metric is
\begin{equation}
ds^2=a^2\eta_{\mu\nu}\mathrm{d}x^\mu \mathrm{d}x^\nu+\mathrm{d}y^2=e^{2A(y)}\eta_{\mu\nu}\mathrm{d}x^\mu \mathrm{d}x^\nu+\mathrm{d}y^2.
\end{equation}

Here we use the covariant approach developed in Ref. \cite{Gong2011}. Recall that Einstein gravity can be written in the Arnowitt--Deser--Misner (ADM) form:
\begin{eqnarray}
ds^2&=&N^2\mathrm{d}y^2+q_{\mu\nu}(\mathrm{d}x^\mu+N^\mu \mathrm{d}y)(\mathrm{d}x^\nu +N^\nu \mathrm{d}y),\\
S&=&\int \mathrm{d}^nx N\sqrt{-q} \left\{ \frac{1}{2\kappa_n} \left[ ^{(n-1)}R - \frac{1}{N^2} q^{\mu\rho}q^{\nu\sigma}\left(E_{\mu\nu} E_{\rho\sigma} - E_{\mu\rho} E_{\nu\sigma} \right) \right] + \mathcal{L}_m \right\} \, ,
\end{eqnarray}
where $ ^{(n-1)}R$ is the $(n-1)$-dimensional curvature scalar constructed from the induced metric $q_{\mu\nu}$, and $E_{\mu\nu}$ is defined by
\begin{equation}
E_{\mu\nu} \equiv \frac{1}{2} \left( \partial_y q_{\mu\nu} - N_{\mu|\nu} - N_{\nu|\mu} \right)
\end{equation}
with a vertical bar denoting a covariant differentiation with respect to $q_{\mu\nu}$. In terms of the ADM variables, the kinetic term $X^{IJ}$ of the scalar fields can be expressed as
\begin{equation}
 X^{IJ}=-{1\over
  2}g^{MN}\partial_M \Phi^I\partial_N \Phi^J=-{1\over
  2N^2}\tilde{\partial}_y\Phi^I\tilde{\partial}_y\Phi^J
  -{1\over 2}q^{\mu\nu}\partial_\mu\Phi^I\partial_\nu\Phi^J \, ,
\end{equation}
where $\tilde{\partial}_y\equiv \partial_y-N^\mu\partial_\mu$.

{\subsection{The action and perturbations}}
In order to obtain the linear perturbed equations we write the scalar fields and the induced metric as follows:
\begin{eqnarray}
  \Phi^I(x,y)&=&\Phi^I_0(y)+\delta \Phi^I(x,y),\\
  q_{\mu\nu}(x,y)&=&q_{(0)\mu\nu}(y)+h_{\mu\nu}(x,y),
\end{eqnarray}
where $q_{(0)\mu\nu}=a^2(y)\eta_{\mu\nu}$ is the background metric.
The perturbation $h_{\mu\nu}$ can be decomposed into the transverse traceless tensor $\bar{h}_{\mu\nu}$, transverse vector  $\bar{v}_{\nu}$ and scalars $\psi,~E$:
\begin{eqnarray}
  h_{\mu\nu}=\bar{h}_{\mu\nu}+2a^2\partial_{(\mu} \bar{v}_{\nu)}
             -2a^2(\eta_{\mu\nu}\psi-\partial_\mu\partial_\nu E),
\end{eqnarray}
where $\eta^{\rho\mu}\partial_\rho\bar h_{\mu\nu}=\eta^{\mu\nu}\bar h_{\mu\nu}=0$, $\eta^{\mu\nu}\partial_\mu\bar v_\nu=0$.
Fluctuations in $N$ and $N_\mu$ are
\begin{eqnarray}
N=1+N_{(1)},\quad N_\mu=\bar{N}_\mu+\partial_\mu B,
\end{eqnarray}
where $\eta^{\mu\nu}\partial_\nu\bar{N}_\mu=0$. Note that perturbations $Q^I=\delta\Phi^I+\frac{u^I}{A'}\psi$ are gauge invariant, where $u^I \equiv \partial_y\Phi^I_0$. In Einstein theory, the  $(n-1)(n-2)/2$ spin-2 tensor modes are physical degrees of freedom, and the vector and scalar modes are non-dynamical since they just give constraint equations. The tensor modes are easy to deal with at the linear level since they are decoupled from the other perturbation modes. The vector modes can be gauged away.
In our model, one can easily check that the tensor modes satisfy
\be
\Box\tilde{h}_{\mu\nu}
+\pt_{z}^{2}\tilde{h}_{\mu\nu}
-\left(\frac{3}{2}\pt_{z}^{2}A
+\frac{9}{4}(\pt_{z}A)^{2}\right)\tilde{h}_{\mu\nu}=0,
\label{tensor shd equation}
\ee
with $\tilde{h}_{\mu\nu}(x^\sigma,z)=\varepsilon_{\mu\nu}(x^\sigma) \psi(z)$, where the conformally flat coordinate $z$ is defined by $\mathrm{d}y=a\mathrm{d}z$. {Note that $\tilde{h}_{\mu\nu}$ is canonical, and we relate $\bar{h}_{\mu\nu}$ with $\tilde{h}_{\mu\nu}e^{-\frac{3}{2}A}=\bar{h}_{\mu\nu}$. The tensor mode} is obviously the same as the standard case, and gives the zero mode solution $\psi_0 (z)=e^{\frac{3}{2}A}$. The localization condition is
\begin{equation}
\int_{-\infty}^{+\infty} |\psi_0 (z)|^2 \mathrm{d}z<\infty.
\end{equation}
For asymptotically $AdS_5$ solutions, like $A(y)=-\log(\cosh(k y))$, this condition can be satisfied, namely the graviton zero mode can be localized, which is necessary but not sufficient for the recovering of four dimensional gravity. In fact, we also need to ensure that the scalar perturbations are stale and the scalar zero modes are not localized on the brane.

For the scalar perturbations, we choose the flat gauge, i.e. $\psi=E=0$. So $Q^I=\delta\Phi^I$. In the flat gauge conditions the perturbations of the brane metric on each slice vanish, which is very useful when we use the ADM formula.
It is worth to emphasize that it is safe for the tensor and scalar perturbations. However, we would lose some information for the vector component. At the non-perturbative level, varying with respect to $N$ and $N^\mu$ we obtain the constraint equations:
\ba
\frac{1}{2} \left[ ^{(n-1)}R +\frac{1}{N^2} q^{\mu\rho}q^{\nu\sigma}\left(E_{\mu\nu} E_{\rho\sigma} - E_{\mu\rho} E_{\nu\sigma} \right) \right] +P+{1\over
  N^2}P_{\langle{IJ}\rangle}\tilde{\partial}_y\Phi^I\tilde{\partial}_y\Phi^J&=&0,\label{scalar-constrain}
  \\
\left[\frac{1}{N}
    (q^{\nu\rho}E_{\mu\rho}-q^{\rho\sigma}E_{\rho\sigma}\delta_\mu^\nu)\right]_{|\nu}
      +\frac{1}{N}P_{\langle{IJ}\rangle}\tilde\partial_y\Phi^I\partial_\mu\Phi^J&=&0,\label{vector-constrain}
\ea
where $ P_{\langle{IJ}\rangle} \equiv \frac{1}{2}\left(\frac{\partial P}{\partial X_{IJ}}+\frac{\partial P}{\partial X_{IJ}}\right)$.

We can extract the background equations of motion from the first-order terms of the action:
\begin{eqnarray}\label{S1}
S_1 &=& \int \mathrm{d}^nx a^{n-1}
      \bigg[
        \Big(-\frac{1}{2}(n-2)(n-1)A'^2 + P_0 + P_{\langle{IJ}\rangle}u^Iu^J \Big) N_{(1)} \nonumber\\
        && +(n-2)A'N^\mu_{(1),\mu}-  P_{\langle{IJ}\rangle}\mathcal{D}_yQ^Iu^J +P_a f_{a;I} Q^I
      \bigg]  ,
\end{eqnarray}
where  $P_0$ is evaluated at the background solution, $\mathcal{D}_y=u^I\mathcal{D}_I$ with $\mathcal{D}_I$ the covariant derivative compatible with the field-space metric $\mathcal{G}_{IJ}$,  and $P_a \equiv \partial P/\partial f_a$.
Taking a variation of (\ref{S1}) with respect to $N_{(1)}$ and $Q^I$,
we obtain
\begin{eqnarray}
(n-2)(n-1)A'^2 &=& 2P_0 +2 P_{\langle{IJ}\rangle}u^Iu^J \, ,\label{Einstein-EOM}
\\
\frac{1}{a^{n-1}}\mathcal{D}_y \left( a^{n-1}P_{\langle{IJ}\rangle}u^J \right) &=&- P_a f_{a;I} \, .\label{scalar-EOM}
\end{eqnarray}
From Eqs. (\ref{Einstein-EOM}) and (\ref{scalar-EOM}) we can derive another Einstein equation:
\begin{equation}
  (n-2)A''=-P_{\langle{IJ}\rangle}u^I u^J.\label{H-primeEOM}
\end{equation}

To obtain the perturbed equations, we can calculate the quadratic order action
\begin{eqnarray}\label{S2}
S_2 = & \int \mathrm{d}^nx\, a^{n-1} \bigg\{ \frac{1}{2} \bigg[ {\mathcal{L}_{\texttt{S}}}+
 P_{\langle{IJ}\rangle\langle{KL}\rangle}\mathcal{D}_yQ^Iu^J\mathcal{D}_yQ^Ku^L
%\nonumber\\
%& \hspace{0.1cm}
- 2P_{\langle{IJ}\rangle{a}}\mathcal{D}_yQ^Iu^J f_{a;K}Q^K
\nonumber\\
& + P_{ab}f_{a;I}f_{b;J}Q^IQ^J \bigg]
 - (n-2)A'N_{(1)}N_{(1),\mu}^{\mu} -
 P_{\langle{IJ}\rangle}N_{(1)}^\mu \partial_\mu Q^Iu^J
 \nonumber\\
&
  + N_{(1)}\Bigl[ P_{\langle{IJ}\rangle}\mathcal{D}_yQ^Iu^J +
 P_af_{a;I}Q^I
 + \left(
 -P_{\langle{IJ}\rangle\langle{KL}\rangle}\mathcal{D}_yQ^Ku^L +
 P_{\langle{IJ}\rangle{a}}f_{a;K}Q^K \right) u^Iu^J \Bigl]
\nonumber\\
& \hspace{0.1cm}
+ \frac{1}{2} N_{(1)}^2
  \Bigl[ (n-1)(n-2)A'^2
 - P_{\langle{IJ}\rangle}u^Iu^J +
 P_{\langle{IJ}\rangle\langle{KL}\rangle}u^I
 u^Ju^Ku^L \Bigl]
  \bigg\} \, ,
\end{eqnarray}
where
\begin{eqnarray}
  {\mathcal{L}_{\texttt{S}}} &=& -P_{\langle{IJ}\rangle}\left[
\mathcal{R}^{(I}{}_{KLR} u^{J)}u^RQ^KQ^L + \mathcal{D}_yQ^I \mathcal{D}_yQ^J\right] \nonumber \\
  &&-  P_{\langle{IJ}\rangle}q^{\mu\nu}\partial_\mu Q^I\partial_\nu Q^J + P_af_{a;IJ}Q^IQ^J.
\end{eqnarray}
Here we have used $N_{(1)\mu}=\partial_\mu B$ for the scalar perturbation and dropped boundary terms.
Varying the quadratic action (\ref{S2}) with respect to $N_{(1)}^\mu$ and~$N_{(1)}$
gives the following constraint equations:
\begin{eqnarray}
\label{lapse_sol}
N_{(1)} &= & \frac{1}{(n-2)A'}P_{\langle{IJ}\rangle}Q^Iu^J\,,
\\
-\frac{(n-2)A'}{a^2} \Box B &= &{\mathcal{P}_{\texttt{S}}}
  -\big(P_{\langle{IJ}\rangle\langle{KL}\rangle}\mathcal{D}_yQ^Ku^L
-P_{\langle{IJ}\rangle{a}}f_{a;K}Q^K \big) u^Iu^J
 \nonumber\\
 &&+N_{(1)} P_{\langle{IJ}\rangle\langle{KL}\rangle}u^I
 u^Ju^Ku^L  ,
\end{eqnarray}
where $\Box=\eta^{\mu\nu}\partial_\mu\partial_\nu$ is the d'Alembert operator on the brane, and
\begin{eqnarray}
 {\mathcal{P}_{\texttt{S}}}=P_{\langle{IJ}\rangle}\mathcal{D}_yQ^Iu^J + P_af_{a;I}Q^I
   +N_{(1)} \big[(n-1)(n-2)A'^2-P_{\langle{IJ}\rangle}u^Iu^J\big].
\end{eqnarray}
The two Lagrange multipliers $N_{(1)}$ and $B$ can be determined by $Q^I$.
Varying the quadratic order action with respect to $Q^I$ and eliminating the Lagrange multipliers, we can obtain the perturbed equations.
Especially, for the simplest case $P=\mathcal{G}_{IJ}X^{IJ}-V$, we have $P_{\langle{IJ}\rangle}=\mathcal{G}_{IJ}$, $f=V$, $P_V=-1$, and so Eqs. (\ref{Einstein-EOM})-(\ref{H-primeEOM}) become
\begin{eqnarray}
  (n-1)(n-2)A'^2&=&\mathcal{G}_{IJ}u^I u^J-2V,\label{phi-cEOM}\\
  \frac{1}{a^{n-1}}\mathcal{D}_y \left( a^{n-1}u^I \right)&=& V^I=
   \mathcal{G}^{IJ}\partial_J V,\label{Einstein-cEOM}\\
  (n-2)A''&=&-u^I u_I.\label{H-prime}
\end{eqnarray}
In this case the quadratic order action can be simplified as
\begin{eqnarray}
  S_2 = & \int \mathrm{d}^nx\, a^{n-1} \bigg\{ \frac{1}{2} \bigg[ -\mathcal{G}_{IJ}\left[
\mathcal{R}^{(I}{}_{KLR} u^{J)}u^RQ^KQ^L + \mathcal{D}_yQ^I \mathcal{D}_yQ^J\right] \nonumber\\
 &-a^{-2}\mathcal{G}_{IJ}\eta^{\mu\nu}\partial_\mu Q^I\partial_\nu Q^J - V_{;IJ}Q^IQ^J\bigg]
+ N_{(1)}\bigl[ \mathcal{G}_{IJ}\mathcal{D}_yQ^Iu^J -
 V_{;I}Q^I \bigl]\nonumber\\
& \hspace{0.1cm} +N_{(1)}^2V- (n-2)A'N_{(1)}N_{(1),\mu}^{\mu} +
 \mathcal{G}_{IJ}N_{(1)}^\mu \partial_\mu Q^Iu^J
 \bigg\} \, .\label{quadratic action}
\end{eqnarray}
The constraint equations read
\begin{eqnarray}
  N_{(1)} &= & \frac{1}{(n-2)A'}Q_Iu^I \label{s-const}\, ,\\
 -\frac{(n-2) A'}{a^2}\Box B &= &u_I \mathcal{D}_yQ^I - V_{;I}Q^I
   -2N_{(1)} V\equiv   {\mathcal{C}}. \label{v-const}
\end{eqnarray}
Varying the action (\ref{quadratic action}) with respect to $Q^I$ and substituting the metric perturbations $N_{(1)}$ and $N^\mu_{(1)}$ by (\ref{s-const}), we obtain
 \begin{eqnarray}
  \frac{1}{a^{n-1}}\mathcal{D}_y(a^{n-1}\mathcal{D}_yQ_I)+\frac{1}{a^2} \Box Q_I-\mathcal{M}_I^J Q_J=0,
  \label{yperturbation}
\end{eqnarray}
where
\begin{eqnarray}
\mathcal{M}_{IJ}&=&V_{;IJ}- \mathcal{R}_{IKJL} u^K u^L+\mathcal{U}_{IJ}, \label{M_IJ}\\
   \mathcal{U}_{IJ}&=&{2\over (n-2)a^{n-1}}\mathcal{D}_y\left({\frac{a^{n-1}}{A'}}u_I u_J\right).
\end{eqnarray}
The second term $\mathcal{R}_{IKJL} u^K u^L$ in Eq.~(\ref{M_IJ}) is a Jacobi term, and the last one $\mathcal{U}_{IJ}$ is an effect of the curved spacetime.

The effective action for $Q^I$ is
\begin{eqnarray}
  S_2 = \frac{1}{2}\int \mathrm{d}^nx\, a^{n-1} \bigg[ -\mathcal{G}_{IJ}
 \mathcal{D}_yQ^I \mathcal{D}_yQ^J -a^{-2}\mathcal{G}_{IJ}\eta^{\mu\nu}\partial_\mu Q^I\partial_\nu Q^J -\mathcal{M}_{IJ}Q^IQ^J\bigg].
\end{eqnarray}
The localized states should satisfy
\begin{equation}
  \int_{-\infty}^{+\infty}\mathrm{d}y a^{n-3}\mathcal{G}_{IJ}Q^IQ^J<+\infty.\label{localized}
\end{equation}
If we define the tetrad fields satisfying
\begin{equation}
  e^i_I e^j_J \delta_{ij}=\mathcal{G}_{IJ},\quad e^i_I e^I_j =\delta^i_j,\quad \mathcal{D}_y e^i_I=0,
\end{equation}
then the fields $Q^I$ can be spanned by the vierbein fields: $Q_I=\sum_i e_I^i Q_i(m^2,y)e^{ip_\mu x^\mu}$ with $\eta^{\mu\nu}p_\mu p_\nu=-m^2$. Now Eq. (\ref{yperturbation}) becomes
\begin{eqnarray}
  \frac{1}{a^{n-3}}\partial_y(a^{n-1}\partial_yQ_i)+m^2 Q_i-a^2\mathcal{M}^j_{i}Q_j=0. \label{yperturbation2}
\end{eqnarray}
where $\mathcal{M}^j_i=\mathcal{M}^J_I e_J^j e_i^I$.
Defining $Q_i= a^{-(n-2)/2}\tilde{Q}_i$, we can rewrite Eq. (\ref{yperturbation2}) as the following coupled Schr\"{o}dinger-like equations:
\begin{eqnarray}
  -\partial_z^2\tilde{Q}_i+
   \left[
     \left(\frac{(n-2)^2}{4}(\partial_zA)^2
      -\frac{(n-2)}{2}\partial_z^2A\right)\delta^j_i+a^2\mathcal{M}^j_i
      \right]
     \tilde{Q}_j =m^2 \tilde{Q}_i. \label{zperturbation}
\end{eqnarray}

\section{Stability and zero modes of the scalar perturbations}\label{zero}

%H.Amann and~P.Quittner~\cite{nodal} have proven the following nodal theorem: The eigenvalue problem of operator $\mathcal{A}$
%\begin{eqnarray}
%    \mathcal{A}Q=\lambda Q,~~Q(b)=0,~~Q\in L_2((0,\infty), \mathbb{R}^n),\label{nod}
%\end{eqnarray}
%the operator $\mathcal{A}$
%\begin{equation}
%  \mathcal{A}Q\equiv-\frac{d}{dy}(C(y)\frac{d}{dy}Q)+(\frac{D(y)}{y^2}+E(y))Q\,,
%\end{equation}
%where~$C, D, E$ are~$(m\times m)$ real symmetric bound matrices, $C(y)\geqslant \alpha>0$ is positive definite, $D(0)$ non-negtive£¬$b>0$ positive, $Q$ is $m$ ·ÖÁ¿ÏòÁ¿£¬³õÖµÎÊÌâ±¾Õ÷Öµ·½³Ì (\ref{nod}) µÄÊø¸¿Ì¬½â($\lambda<0$)ÊýÄ¿µÈÓÚÆë´Î·½³Ì~~$\mathcal{A}Q=0, Q(b)=0$ µÄ¶ÀÁ¢½â×é³ÉµÄ½â¾ØÕóÐÐÁÐʽµÄÁã¸ù¸öÊý¡£µ±~$D(0)=0$£¬¿ÉÒÔÉè~ $b=0$¡£

 Contracting Eq. (\ref{yperturbation}) with $u_I$, we have
 \begin{eqnarray}
%  &&\frac{u^I}{a^{n-1}}\mathcal{D}_y(a^{n-1}\mathcal{D}_yQ_I)-s u^I V_{;IJ}Q^J-u^I(N_{(1)}'u_I+2sN_{(1)}V_{,I}+\frac{1}{a^2}u_I\Box B)\nonumber\\
%  &=&
  \frac{A'}{a^{n-1}}\left(\frac{a^{n-1}\mathcal{C}}{A'}\right)'
  =-\frac{m^2}{a^2}u^IQ_I.\label{ani-cons}
\end{eqnarray}
Using the definition of $\mathcal{C}$ in (\ref{v-const}) and the constraint equation for $N_{(1)}$ in (\ref{s-const}), we can obtain the anisotropic constraint from Eq. (\ref{ani-cons}):
\begin{eqnarray}
\frac{1}{a^{n-3}}(a^{n-3}B)'=N_{(1)}\label{ani-cons-2}.
\end{eqnarray}
This constraint can be derived from the linearised Einstein equations without choosing the flat gauge, and it is not independent. For $m^2=0$, it is the Eq.
(\ref{ani-cons-2}) that fixes $B$, but not the constraint (\ref{v-const}).
This means that we do not lose any equations, which does happen for the vector perturbation if a gauge condition is chosen before calculating the quadratic order action.

By similar process we can prove that it still holds for the general case (\ref{S2}).
For a single-scalar braneworld model its perturbation equation can be factorized, and can be rewritten as a supersymmetric equation after redefining perturbed field. So the braneworld constructed by a single scalar field is stable if there exists no ghost ($P_X>0$). This conclusion can be generalized to the case of a K field.

In order to recover normal gravitational potential on the brane, we require that no massless and tachyon modes localize on the brane. For a general multi-field solution, as far as we know, there is no available result analogous to that of the tensor modes. Fortunately, there are some mathematical results which can be used to deal with the coupled Schr\"{o}dinger equation (\ref{zperturbation}).  According to the nodal theorem of Schr\"{o}dinger equation \cite{nodal,George2011,Aybat2010,Gherghetta:2011rr},
one could define a solution matrix with all the zero mode solutions, then the number of bound states (in our case they are tachyons) equals  the number of zero roots of the determinant. In other words, now we can determine the stability of braneworld solutions using the solutions of the zero modes.

For multiple scalar cases, we can separate the field space into the background trajectory direction and its orthogonal space, and the corresponding perturbed modes are $Q_\sigma$ and $\vec{Q}_s$, respectively. The divergent term $\mathcal{U}_{IJ}$ only directly affects the modes $Q_\sigma$, while the Jacobian term $\mathcal{R}_{IKJL} u^K u^L$ only acts non-trivially on the modes $\vec{Q}_s$.

Since the potential $\mathcal{U}_{IJ}$ is singular near the position of the brane, we should choose proper initial conditions at $y=0$. The first initial conditions are $Q^I(0)=0$, which require $Q_\sigma'(0)=0$. So $\vec{Q}_s'(0)$ and $Q_\sigma''(0)$ are initial parameters. This technique has been widely used in cosmology \cite{Wands1996,Gordon2000,Gordon2013} and has been proved to be very fruitful. All of the independent $\vec{Q}_s'(0)$ and $Q_\sigma''(0)$ give all possible states satisfying $Q^I(0)=0$. These solutions form a solution matrix. According to the nodal theorem for coupled systems of Schr\"{o}dinger equations \cite{nodal}, the number of zeros of its determinant equals the number of the tachyon fields.

There are other possible initial conditions. Apart from a singular solution, we can choose $\vec{Q}_s(0)=\vec{c}$, $\vec{Q}_s'(0)=\vec{0}$. The singular solution can also satisfy this condition. These solutions form another solution matrix. For this section no mathematically rigorous theorem exists. However, there are some hints which suggest that a similar result is available for this section \cite{George2011, Aybat2010}.

{For a double-field model, we can define $\sigma^I=\frac{u^I}{u}, u\equiv |u^I|=\sqrt{-(n-2)A''}$, and the adiabatic mode $Q_\sigma=\sigma^IQ_I$. In general another vector $s^I$ can be a unit vector orthogonal to $\sigma^I$. If $\omega\equiv |\mathcal{D}_y\sigma_I|\neq0$, we can choose $s^I=\frac{\mathcal{D}_y\sigma^I}{|\mathcal{D}_y\sigma^I|}$, and the entropy mode is then $Q_s=s^I Q_I$.}
The localizable condition (\ref{localized}) becomes
\begin{equation}
  \int_{-\infty}^{+\infty}\mathrm{d}y a^{n-3}(\vec{Q}_s^2+Q_\sigma^2)<\infty.
\end{equation}

%Zero modes $m^2=0$ are important for the model: They can be used to analyze the problem of stability.
Now we consider the $m^2=0$ limit.
The massless modes obey the following equations and constraints:
\begin{eqnarray}
  \frac{1}{a^{n-1}}\mathcal{D}_y(a^{n-1}\mathcal{D}_yQ_I)-V_{;IJ}Q^J+\mathcal{R}_{IKJL} u^K u^L Q^J
     -\mathcal{U}_I^JQ_J&=&0,\label{scalar0}\\
  \mathcal{C} =u_I \mathcal{D}_yQ^I - V_{;I}Q^I-\frac{2}{(n-2)A'}Q_Iu^IV&=&0.\label{C0}
\end{eqnarray}
%For the massless case $p^2=0$, Eq. (\ref{ani-cons}) gives
%\begin{equation}
%\mathcal{C}=c_1A'e^{-(n-1)A}.
%\end{equation}
The constraint (\ref{C0}) is compatible with the perturbed equations (\ref{yperturbation}).

We can write the singular massless mode explicitly:
\begin{equation}
  Q_I=u_I/A'.\label{uni-sol}
\end{equation}
%So there are 2 independent solutions within constraint, or 3 without constraint.
We take a double-field model for example.
 Then the constraints (\ref{s-const}) and (\ref{v-const}) become
\begin{eqnarray}
 N_{(1)}&=&\frac{1}{(n-2)A'}u Q_\sigma, \\
 \mathcal{C}&=&u(\partial_yQ_\sigma-\frac{u'}{u}Q_\sigma+\frac{A''}{A'}Q_\sigma-2\omega Q_s)=-\frac{(n-2)A' m^2}{a^2}B.
\end{eqnarray}
%Contracting with $s^I$ gives
%\begin{eqnarray}
%\partial^2_y Q_s+(n-1)A'\partial_y Q_s-(V_{ss}-\mathcal{R}_{IKJL}u^K u^L s^I s^J-3\omega^2-\frac{m^2}{a^2})Q_s+\frac{2\omega}{u}\mathcal{C}=0.
%\end{eqnarray}
%where $V_{ss}=V_{;IJ}s^I s^J$.
%
%
%%Zero modes satisfy the following equations and constraint:
%%\begin{eqnarray}
%%  \frac{1}{a^{n-1}}\mathcal{D}_y(a^{n-1}\mathcal{D}_yQ_I)-V_{;IJ}Q^J+\mathcal{R}_{IKJL} u^K u^L
%%     -\mathcal{U}_I^JQ_J&=&0,\label{scalar0}\\
%%  \mathcal{C} =u_I \mathcal{D}_yQ^I - V_{;I}Q^I-\frac{2}{(n-2)A'}Q_Iu^IV&=&0.\label{C0}
%%\end{eqnarray}
%
%
%
%
%One can write {\color{red}the constraints (\ref{s-const}) and (\ref{v-const})} in {\color{red}terms} of $Q_\sigma$ and $Q_s$
%\begin{eqnarray}
%\mathcal{C}&=&u(\partial_yQ_\sigma-\frac{u'}{u}Q_\sigma+\frac{A''}{A'}Q_\sigma-2\omega Q_s)=-\frac{(n-2)A'm^2}{a^2}B,\\
%N_{(1)}&=&\frac{1}{(n-2)A'}u Q_\sigma.
%\end{eqnarray}
Contracting Eq. (\ref{yperturbation}) with $s^I$ yields
\begin{eqnarray}
%&&\partial^2_y Q_s+(n-1)A'\partial_y Q_s-(\omega^2+V_{ss}-\mathcal{R}_{IKJL}u^K u^L s^I s^J-\frac{m^2}{a^2})Q_s
%+2\omega(\partial_yQ_\sigma-\frac{u'}{u}Q_\sigma+\frac{A''}{A'}Q_\sigma)\nonumber\\
%&=&
\partial^2_y Q_s+(n-1)A'\partial_y Q_s-\left( V_{ss}-\mathcal{R}_{IKJL}u^K u^L s^I s^J-3\omega^2-\frac{m^2}{a^2}\right)Q_s+\frac{2\omega}{u}\mathcal{C}=0,
\end{eqnarray}
with $V_{ss}=V_{;IJ}s^I s^J$. When $\omega=0$, $Q_\sigma$ and $Q_s$ decouple with each other. Then the mode $Q_\sigma$ is stable and the stability of the mode $Q_s$ can be determined by a single Schr\"{o}dinger equation.

For the zero modes of the double-scalar model, we rewrite the constraint and the perturbed equations as follows
\begin{eqnarray}
\partial^2_y Q_s+(n-1)A'\partial_y Q_s-(V_{ss}-\mathcal{R}_{IKJL}u^K u^L s^I s^J-3\omega^2)Q_s&=&0,\label{homoQs} \\
\partial_yQ_\sigma-\frac{u'}{u}Q_\sigma+\frac{A''}{A'}Q_\sigma-2\omega Q_s&=&0.\label{homo}
\end{eqnarray}
In the massless limit the modes $Q_\sigma$ and $Q_s$ are decoupled.

Regardless of the constraint (\ref{C0}), Eq. (\ref{ani-cons}) for the massless case $p^2=0$ yields
\begin{equation}
\mathcal{C}=c_1A'e^{-(n-1)A}.\label{ani-cons-non}
\end{equation}
$c_1$ is a constant. It is clear that the constraint (\ref{C0}) requires $c_1=0$. The physical massless modes should satisfy this constraint. However, other modes have no corresponding constraint. Therefore, in order to analyze the stability of this system, we must loosen this constraint. It gives an inhomogeneous equation. We can write the equation of the zero mode in the following form:
\begin{equation}
\partial^2_y Q_s+(n-1)A'\partial_y Q_s-\left(V_{ss}-\mathcal{R}_{IKJL}u^K u^L s^I s^J-3\omega^2\right)Q_s=-\frac{2\omega}{u}\mathcal{C}. \label{non-homo}
\end{equation}
If $c_1=0$, then it is homogeneous, and its initial conditions $Q_s(0)=0$, $Q_s'(0)=1$ or $Q_s(0)=1$, $Q_s'(0)=0$ can give possible physical modes; If $c_1=1$, the homogeneous initial condition $Q_s(0)=0$, $Q_s'(0)=0$ gives an additional solution.

From Eq.~(\ref{homo}) it is easy to obtain
\begin{equation}
  \frac{A'}{u}Q_\sigma=\int \mathrm{d}y \frac{A'}{u}\left(\frac{\mathcal{C}}{u}+2\omega Q_s\right).
   \label{connect-us}
\end{equation}
If~$\mathcal{C}=Q_s=0$, we get the universal solution (\ref{uni-sol}).
Redefining the perturbed fields $\tilde{Q}_s=a^{\frac{n-1}{2}}Q_s$, Eq. (\ref{homoQs}) can be written as a Schr\"{o}dinger-like equation
\begin{eqnarray}
  -\partial_y^2\tilde{Q}_s+V_f \tilde{Q}_s=0,\label{Schrodinger0}
  \end{eqnarray}
where the potential $V_f$ is given by
\begin{equation}\label{perturbed eq}
  V_f=V_{ss}-\mathcal{R}_{IKJL}u^K u^L s^I s^J-3\omega^2+\frac{(n-1)^2}{4}A'^2+\frac{n-1}{2}A''\,.
\end{equation}
From the above potential we can know whether the zero mode can be localized. According to the conjectures proposed in Ref.~\cite{George2011}, if there is no localized state in Eq. (\ref{Schrodinger0}), then the zero mode cannot have zero point.

We should mention a very special category:
\begin{eqnarray}
  V&=&\frac{(n-2)^2}{2}\mathcal{G}^{IJ}W_{,I}W_{,J}-\frac{(n-1)(n-2)}{2}W^2,\\
  A'&=&-W,~~u^I=(n-2)\mathcal{G}^{IJ} W_{,J},
\end{eqnarray}
where $W$ is a superpotential. For a superpotential solution, one finds that its linearized perturbed equations can be written as follows \cite{BHM2005}
\begin{eqnarray}
 \left(-\delta^I_J\mathcal{D}_y-Z^I_J+(n-1)\delta^I_J W\right)
 \left(\delta^J_K\mathcal{D}_y-Z^J_K \right) Q^K
  =\frac{m^2}{a^2}Q^I
\end{eqnarray}
with
\begin{equation}
  Z^I_J=(n-2)\left(W^I_{;J}-\frac{W^I W_J}{W}\right).
\end{equation}
It is supersymmetric:
\begin{eqnarray}
  \int dy e^{(n-3)A}m^2Q^IQ_I
  &=&\int dy a^{n-1}Q_I
  \left(-\delta^I_J\mathcal{D}_y-Z^I_J+(n-1)W\delta^I_J \right)
  \left(\delta^J_K\mathcal{D}_y-Z^J_K\right) Q^K\nonumber\\
  &=&\int dy a^{n-1}|\mathcal{D}_y Q^I-Z^I_J Q^J|^2\geq0.\label{super-perb}
\end{eqnarray}
So~$m^2\geq0$, the model is stable. From Eq. (\ref{super-perb}), the zero modes satisfy
\begin{equation}\label{super-zero}
  \mathcal{D}_y Q^I-Z^I_J Q^J=0.
\end{equation}
Contracting with $u_I$, we obtain the constraint (\ref{ani-cons-non}). For the zero modes of the superpotential solutions, the constraint (\ref{ani-cons-non}) is a conclusion of the perturbation equations. Whether the massless modes can be localized is totally determined by the asymptotic behavior of the background solution. If $W$ reaches its minimum, there always exist some bound massless states. For the double-scalar superpotential case, from Eq.~(\ref{super-zero}) we derive
\begin{eqnarray}
  Q_s'-(n-2)W_{ss}Q_s=0,
\end{eqnarray}
where $W_{ss}=W_{IJ}s^I s^J$.
$Q_s$ and $Q_\sigma$ can be given by
\begin{eqnarray}
Q_s&=&e^{(n-2)\int dy W_{ss}},\\
Q_\sigma &=&\frac{\sqrt{W'}}{W}\int dy \frac{\omega W}{\sqrt{W'}} Q_s.
\end{eqnarray}
{This solution can be normalized.} It means that there exists a remnant massless scalar field on the brane. For the case of five dimensions, this result is conflicted with observations and is not acceptable \cite{George2011}.

\section{Stability of $f(R)$ braneworld }\label{secfR}
{\subsection{The Einstein frame formalism}}
The above method can be generalized to analyze the stability of other gravity theories. The simplest ones are non-minimal coupling gravity theories. In this section, we deal with the $f(R)$ gravity theory which can be treated as a special non-minimal coupling gravity theory.

The action of the multi-field metric $f(R)$ gravity reads
\begin{equation}\label{fR1}
S=\int \mathrm{d}^nx\sqrt{-g}
 \left(\frac{1}{2}f(R)-\frac{1}{2}g^{MN}\partial_M\vec{\phi}\partial_N\vec{\phi}-V(\phi)\right).
\end{equation}
Introducing an auxiliary field $\chi$, the action (\ref{fR1}) can be written as
\begin{eqnarray}
 S=\int \mathrm{d}^nx\sqrt{-g}
  \left[\frac{1}{2}f_R(\chi)R-\frac{1}{2}g^{MN}\partial_M\vec{\phi}\partial_N\vec{\phi}
  -V(\phi)-\frac{1}{2}(\chi f_R-f)\right],
\end{eqnarray}
where~$f_R\equiv df/dR$. The equation of motion for the scalar field $\chi$ is $\chi=R$. Making a conformal transformation \
$\tilde{g}_{MN} = e^{2\varphi} g_{MN} $ with $\varphi=\frac{1}{n-2}\ln f_R(\chi) $,
one can obtain
\begin{eqnarray}
S&=&\int \mathrm{d}^nx\sqrt{-\tilde{g}}
  \left\{\frac{1}{2}\left[\tilde{R}-(n-1)(n-2)\tilde{g}^{MN}\partial_M\varphi\partial_N\varphi
    \right]+\tilde{\mathcal{L}}_m \right\},\\
\tilde{\mathcal{L}}_m&=&
  -\frac{1}{2f_R(\chi)}\tilde{g}^{MN}\partial_M\vec{\phi}\partial_N\vec{\phi}
  -f_R^{-\frac{n}{n-2}} \left[V(\phi)+\frac{1}{2}(\chi f_R-f)\right].
\end{eqnarray}
In this frame the gravity action is the Einstein--Hilbert form; however, the matter fields are non-minimally coupled. For convenience we define a new scalar field:
\begin{eqnarray}
\zeta=\sqrt{\frac{n-1}{n-2}}\ln f_R(\chi)=\frac{1}{2K}\ln f_R(\chi),
\end{eqnarray}
where $K=\sqrt{\frac{n-2}{4(n-1)}}$ is a positive constant.
The metric of the field space ($\Phi^I=(\zeta,\vec{\phi})$) is
\begin{eqnarray}
\mathcal{G}_{IJ}d\Phi^I d\Phi^J=d\zeta^2+e^{-2K\zeta}ds^2_{\vec{\phi}}.
\end{eqnarray}
It is a warped geometry and the warped factor is $e^{-2K\zeta}$. For a non-trivial $f(R)$ gravity, $ f_{RR}\neq 0$, the corresponding Riemann curvature does not vanish. In particular, if the original scalar fields are minimally coupled, the resulting field space is an AdS space.

Different~$f(R)$ theories give different potentials:
\begin{eqnarray}
  \tilde{V}(\zeta,\vec{\phi})=
  e^{-\frac{2n}{n-2}K\zeta} \left[V(\phi)+\frac{1}{2}\left(e^{2K\zeta}\chi(\zeta)-f(\chi(\zeta))\right)\right].
\end{eqnarray}
We should transform the physical coordinates to the conformal ones, and $e^{2A}$ should also be replaced by $e^{2(A+\varphi)}$.
In Ref.~\cite{LiuZhongZhaoLi2011}, a five-dimensional flat braneworld solution for $f(R)=R+\gamma R^2$ was obtained:
\begin{eqnarray}
ds^2&=&e^{2A}\eta_{\mu\nu}dx^\mu dx^\nu+dy^2,\\
V(\phi)&=&\lambda^{(5)}({\phi^2}-v^2)^2+\Lambda_5,\\
%
%\lambda^{(5)}&=&\frac{3}{784\gamma},~~~v=7\sqrt{\frac{3}{29}},~~
%\Lambda_5=-\frac{477}{6728}\frac{1}{\gamma},\\
%
\phi(y)&=&v\tanh {(k y)},\\
e^{A(y)}&=&\text{sech} (k y),
\end{eqnarray}
where
\begin{eqnarray}
\lambda^{(5)}=\frac{3}{784\gamma},~~~v=7\sqrt{\frac{3}{29}},~~
\Lambda_5=-\frac{477}{6728}\frac{1}{\gamma},~~~k=\sqrt{\frac{3}{232\gamma}}.
\end{eqnarray}
In this model $\gamma$ is the only one parameter.

\begin{figure}[htb]
\begin{center}
\subfigure[The perturbation potential]{
\includegraphics[width=6.0cm]{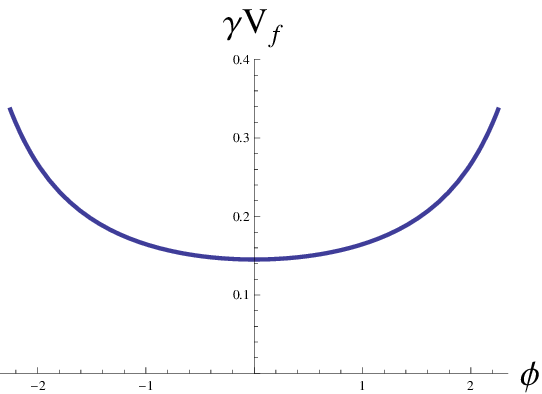}\label{fR-Vfa}}
\subfigure[The even zero mode]{
\includegraphics[width=6.0cm]{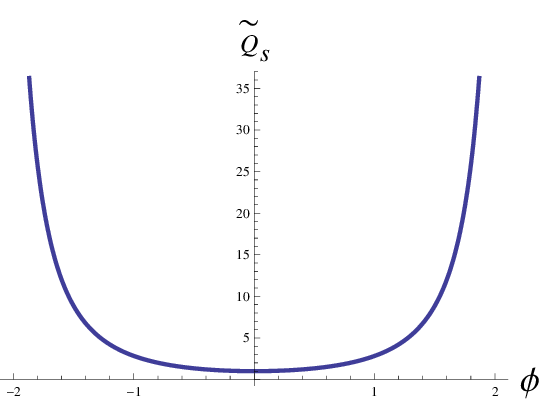}\label{fR-Vfb}}
\end{center}
 \caption{The perturbation potential~$V_f$ and the even zero mode solution $\tilde{Q}_s$ for~$f(R)=R+\gamma R^2$.}
 \label{fR-Vf}
\end{figure}

Converting to double scalar fields in the Einstein frame, the new scalar field and the potential are
\begin{eqnarray}
\zeta&=&\frac{2}{\sqrt{3}}\ln(2\gamma\chi+1),\\
  \tilde{V}&=&e^{-10K\zeta/3}
   \left[\lambda^{(5)}(\phi^2-v^2)^2+\Lambda_5+\frac{(e^{2 K\zeta}-1)^2}{8\gamma}\right].
\end{eqnarray}
where $K=\sqrt{3}/4$.
In the conformal coordinates, we have
\begin{eqnarray}
kz&=&\sinh ky,\\
\phi&=&v\frac{kz}{\sqrt{1+k^2z^2}},~~e^{2A}=\frac{1}{1+k^2z^2},\\
 \zeta&=&\frac{2}{\sqrt{3}}\ln(2\gamma\chi+1)=\frac{2}{\sqrt{3}}
  \ln\left(\frac{14}{29}+\frac{21}{29}\frac{1}{1+k^2z^2}\right).
\end{eqnarray}
In this model, the perturbed equation can be independent of the parameter $\gamma$, and
the configuration of the perturbation potential $V_f$ in Eq.~(\ref{perturbed eq}) under the $\phi$ coordinate is shown in Fig. \ref{fR-Vfa}.
From the plot, we can conclude that there is no localized scalar mode.
Figure~\ref{fR-Vfb} gives the numerical zero mode $\tilde{Q}_s$, from which it can be seen that there is no zero point in the $\phi>0$ region.

\begin{figure}[htb]
\begin{center}
\subfigure[The determinant of the solution matrix]{\label{fR-det1}
\includegraphics[width=6.0cm]{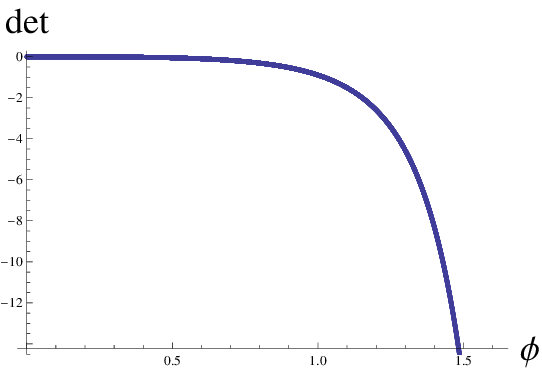}}
\subfigure[Zoom in of Fig.~\ref{fR-det1} around $\phi=0$]{\label{fR-det2}
\includegraphics[width=6.0cm]{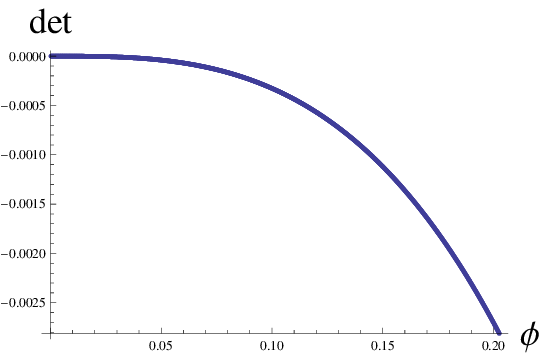}}
\end{center}
 \caption{The determinant of the solution matrix for~$f(R)=R+\gamma R^2$. }
 \label{fR-det}
\end{figure}

\begin{figure}[htb]
\begin{center}
\includegraphics[width=6.0cm]{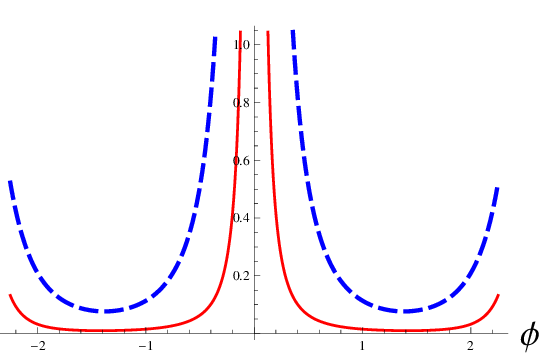}
\end{center}
 \caption{The determinant of the matrix~$\gamma\mathcal{M}_{IJ}$ (the solid red line) and $\gamma\mathcal{M}_{11}$ (the dashed blue line) for~$f(R)=R+\gamma R^2$.}
 \label{fR-zero-M}
\end{figure}

{\subsection{Stability of $f(R)$ brane}}
The odd solution and the homogeneous solution of the inhomogeneous equations form a solution matrix. As is shown in Fig.~\ref{fR-det}, the numerical determinant of the solution matrix has no zero point. So there is no tachyon and the model is stable. We can also see that the massless mode cannot be localized.
Actually, we find that the matrix $\mathcal{M}_{IJ}$ is positive definite for this special solution. The determinant of the matrix~$\gamma\mathcal{M}_{IJ}$ and the component $\gamma\mathcal{M}_{11}$ are plotted in Fig.~\ref{fR-zero-M}.
For all eigenvalues $m^2$, we have
\begin{eqnarray}
  \int dy m^2a^{2}\mathcal{G}_{IJ}Q^I Q^J
  =\int dy a^{4}(\mathcal{G}_{IJ}\partial_yQ^I \partial_yQ^J+\mathcal{M}_{IJ}Q^I Q^J)>0.
\end{eqnarray}
So $m^2>0$.
It agrees with the previous numerical result. So the background solution is stable and the zero modes cannot be localized on the brane.

\section{Conclusions}\label{conclusion}

In this paper, we have investigated the stability of the tensor and scalar perturbations for flat braneworld models constructed by non-minimally coupled multi-scalar fields in the Einstein frame.

Firstly, we studied the stability of the tensor perturbation and find that its dynamical equation can be written as a supersymmetric Schr\"{o}dinger equation, so it is stable at linear level. It was also shown that the tensor zero mode can be localized on the brane if the bulk geometry is asymptotically $AdS_5$. This is the same as the case of a single-field brane model.

Secondly, we presented a systematic covariant approach in field space to deal with the stability problem for the scalar perturbations. The covariant quadratic order action and the corresponding first-order perturbed equations were derived. But these equations cannot be used to analyze the stability of the scalar perturbations. Thus, by introducing the orthonormal bases in field space and making the Kaluza--Klein decomposition, we showed that the Kaluza--Klein modes of the scalar perturbations satisfy a set of coupled Schr\"{o}dinger-like equations. It was shown that these equations for the scalar perturbations are complete. Thus, according to the nodal theorem for the coupled Schr\"{o}dinger equations, we can analyze the stability of the scalar perturbations and localization of the scalar zero modes. For {brane} models constructed with superpotential method, it was shown that the scalar perturbations are stable, while the scalar zero modes are normalizable and can be localized on the brane. Such localized scalar zero modes will result in an unacceptable fifth force on the brane.

Lastly, we applied this approach to the $f(R)$ gravity coupled with one or more scalar fields. By introducing an auxiliary field and a conformal transformation, the $f(R)$ theory was changed to the Einstein frame with non-minimally coupled multi-scalar fields. This procedure leads to a warped field-space geometry. Especially, we tested a particular $f(R)$-brane solution given in Ref.~\cite{LiuZhongZhaoLi2011} and found that the scalar perturbations are stable and there is no normalizable scalar zero mode. Besides, it has been shown that the tensor zero mode of the perturbations can be localized on the $f(R)$ brane~\cite{LiuZhongZhaoLi2011}. Therefore, we can conclude that the $f(R)$ brane model is stable under the linear tensor and scalar perturbations and the four-dimensional Newtonian potential on the brane can be recovered.

We can also analyze scalar perturbations of other modified gravity theories by quadratic order action. Eddington-inspired Born--Infeld gravity is an example \cite{Banados2010,yangkedewenzhang2012,FuLiu2014}, which can be rewritten as a bimetric-like theory \cite{Delsate:2012ky, Lagos:2013aua}. We leave this for future work.

\section*{Acknowledgements}

This work was supported by the National Natural Science Foundation of China (Grants nos. 11522541, and 11375075), and the Fundamental Research Funds for the Central Universities (Grant no. lzujbky-2016-k04).

\end{document}